\documentclass[a4paper,11pt]{article}
\pdfoutput=1 % if your are submitting a pdflatex (i.e. if you have
\usepackage{jheppub} % for details on the use of the package, please
\usepackage[T1]{fontenc} % if needed
\usepackage{amssymb}
\usepackage{amsmath}
\usepackage{color}
\usepackage{xspace}
\usepackage{appendix}
\usepackage{float}

\title{\boldmath    Inclusive heavy flavor jet production  with semi-inclusive jet functions:  from proton  to  heavy-ion collisions}

\author[a]{Hai Tao Li }
\author[a]{and Ivan Vitev }
\affiliation[a]{Los Alamos National Laboratory, Theoretical Division, Los Alamos, NM, 87545, USA}

\emailAdd{haitaoli@lanl.gov}
\emailAdd{ivitev@lanl.gov}

\abstract{
The past several years have witnessed important developments in the QCD theory of jet production and jet substructure in hadronic collisions.
In the framework of soft-collinear effective theory, semi-inclusive jet functions and semi-inclusive fragmenting jet functions have allowed 
us to combine higher order calculations with
resummation of potentially large logarithms of the jet radius,   $\ln R$. Very recently, the semi-inclusive jet functions for partons fragmenting
into heavy flavor jets were computed by Dai, Kim and Leibovich. In this paper we show how the formalism can be extended to 
c-jet and b-jet production in heavy ion collisions.  The semi-inclusive jet functions 
for heavy flavor jets in a QCD medium are evaluated up to the next-to-leading order in $\alpha_s$ and  first order in opacity. For 
phenomenological  applications,  we also consider the inclusion of the cold nuclear matter effects and the jet energy dissipation due 
to collisional interactions in matter.  We present the numerical predictions for the cross sections and the corresponding nuclear modification 
factors  in proton-nucleus and nucleus-nucleus collisions and compare our results to  data from the Large Hadron Collider. 
}

\preprint{LA-UR-19-20952}

\begin{document} 
\maketitle
\flushbottom

\section{Introduction}
\label{sect:int}

Jet production is one of the cornerstone perturbative Quantum Chromodynamics (QCD)  processes in hadronic collisions~\cite{Sterman:1977wj}.  It is characterized by large cross sections and has, thus, been measured with unprecedented precision in comparison to other high energy processes. Observables related to jets have served as precision tests of QCD and as tools to search for new physics. In this work, we restrict our discussion on the inclusive heavy flavor-tagged  jet production in proton and heavy-ion collisions, which has been measured at the Large Hadron Collider (LHC)~\cite{Chatrchyan:2012dk,Chatrchyan:2013exa,Khachatryan:2015sva, Sirunyan:2016fcs} and will be measured at the Relativistic Heavy Ion Collider (RHIC)~\cite{Adare:2015kwa} in the near future. 

Inclusive jet production is a multiscale problem in both  proton-proton (p+p) and heavy ion  (p+A, A+A) collisions. The differential jet cross section versus  transverse momentum $p_T$ and  rapidity $\eta$ in hadronic collisions is factorized as the convolution of the parton distribution functions (PDFs), the hard kernels, and the semi-inclusive jet functions (SiJFs),  or the  fragmentation functions to jet ~\cite{Kaufmann:2015hma}:
\begin{align}\label{eq:main}
    \frac{d\sigma_{pp\to J +X}}{dp_T d\eta} =& \frac{2p_T}{s}\sum_{a,b,c} \int_{x_a^{\rm min}}^1 \frac{dx_a}{x_a} f_a(x_a,\mu) \int_{x_b^{\rm min}}^1 \frac{dx_b}{x_b} f_b(x_b,\mu)
    \nonumber \\ &\times
    \int_{z_{\rm min}}^1 
     \frac{dz_c}{z_c^2}
    \frac{d\hat{\sigma}_{ab\to c}(\hat{s}, p_T/z_c, \hat{\eta}, \mu)}{dv dz} J_{ J/c}(z_c,  w_J \tan(R^\prime/2), m_Q, \mu)~,
\end{align}
where $f_{a,b}$ denotes the usual PDF for parton $a$ or $b$, $d\hat{\sigma}_{ab\to c}(\hat{s}, p_T, \hat{\eta}, \mu)/dv dz$ represents the hard function for the sub-process $a b \to c $, and  $J_{ J/c}$ is the jet function describing the probability of a parton $c$ with transverse momentum $p_T/z_c$ to fragment into a jet $J$ with $p_T$. The variables $\hat{s}$ and $\hat{\eta}$ are the partonic center-of-mass energy and parton rapidity. The SiJFs depend on the momentum fraction $z_c$ taken by the jet, the jet energy $w_J$,  $R^\prime=R/\cosh(\eta)$ from  the  jet reconstruction algorithm,  and the quark mass $m_Q$. The $\ln R$ resummation for  jet production can be achieved by evolving  the  SiJFs from the jet scale $\mu_J$ to the factorization scale $\mu$. The renormalization group (RG)  equations are the usual
 time-like DGLAP evolution equations. The corresponding expressions for the hard part  can be found in Refs.~\cite{Aversa:1988vb,Jager:2002xm}. The SiJFs have been calculated up to next-to-leading order (NLO), using soft-collinear effective theory (SCET)~\cite{Bauer:2000ew,Bauer:2000yr,Bauer:2001ct,Bauer:2001yt,Beneke:2002ph} techniques,  for light jet~\cite{Kang:2016mcy,Dai:2016hzf} and very recently for heavy flavor jets~\cite{Dai:2018ywt}.  
The power corrections in proton collisions to the factorized cross section in Eq.~(\ref{eq:main}) are of order $\Lambda_{\rm QCD}^2/p_T^2$ and jet radius $R^{\prime 2}$. 
 This factorization formula has been used to predict the inclusive  and heavy flavor-tagged jet cross sections~\cite{Dasgupta:2014yra,Dasgupta:2016bnd, Kaufmann:2015hma, Kang:2016mcy,Dai:2016hzf, Dai:2018ywt}. The first application to heavy-ion collisions for light jet was given by Ref.~\cite{Kang:2017frl}.  In this paper we will extend the formalism to the case of inclusive c-jet and b-jet production in reactions with nuclei at ultra-relativistic energies.  We will include  all leading contributions consistent with the power counting of SCET$_{\rm (M,) G}$~\cite{Ovanesyan:2011xy,Kang:2016ofv}. The medium modifications considered here are accurate up to $\mathcal{O}(\alpha_s)$ in QCD,  to first order in the opacity of nuclear matter, and $\mathcal{O}(\mu^2/p_T^2)$ in SCET$_{\rm (M,) G}$ where $\mu$ is the Debye screening mass.

In heavy-ion collisions, one expects the formation of a new deconfined state of matter, known as the quark-gluon plasma  (QGP).  Energetic parton propagation and  shower formation in this strongly interacting matter alter light particle, heavy flavor, and jet observables  in heavy-ion relative to proton collisions -  a phenomenon  known as the jet quenching effect. Studies of jet quenching, especially for the heavy flavor-tagged jets,  can reveal the fundamental thermodynamic and transport properties of the QGP. For an experimental perspective on this issue see~\cite{Connors:2017ptx} and references therein.  It is well-understood that the jet quenching effects depend on the flavor of the fragmenting parton which, together with  heavy quark mass effects, can be studied by light jet and heavy flavor tagged jet (c-jet and b-jet) observables in heavy ion collisions. Mass effects are expected to play   a
significant role in the small and intermediate transverse momentum regions and are, of course, most pronounced for b-jets~\cite{Huang:2013vaa,Li:2017wwc,Kang:2018wrs}. 
They can also change the relative importance of radiative and collisional processes in the QGP for the modification of jet cross sections and jet substructure.

Heavy flavor studies are, therefore,  central to high-energy nuclear physics since they provide new avenues to explore QCD in  strongly-interacting matter and new diagnostics of its
transport properties~\cite{Andronic:2015wma,Rapp:2018qla,Cao:2018ews}.  These efforts have been directed toward open heavy flavor, namely D-meson and B-meson production, and 
quarkonia. Investigation of  heavy-flavor tagged jet production in heavy ion collisions has been somewhat limited thus far.  Inclusive b-jets were first studied using the 
traditional energy loss approach~\cite{Huang:2013vaa} and later with the means of a partonic transport model~\cite{Senzel:2016qau}.  Strategies to suppress the contribution from gluon splitting and enhance the fraction of prompt b-jets  via energetic photon or  B-meson coincident measurements opposite to the b-jet  were performed in Ref.~\cite{Huang:2015mva}. This was generalized to back-to-back b-jets~\cite{Dai:2018mhw,Kang:2018wrs}, moreover  heavy flavor dijet  mass distributions  in heavy ion collisions were also
computed~\cite{Kang:2018wrs}. 
The derivation of full in-medium splitting functions for light and heavy partons~\cite{Ovanesyan:2011kn,Kang:2016ofv,Sievert:2018imd} allows us to bridge the gap between  high energy and heavy
ion theory of jet and heavy flavor production in hadronic and nuclear collisions. Their first application to b-tagged jet substructure~\cite{Li:2017wwc} has already  produced novel results, namely a unique inversion of the mass hierarchy of jet quenching effects as manifested in the stronger modification of  the b-jet momentum sharing distributions  
in comparison to the ones for light jets. These effects are driven by the heavy quark mass and are measurable at the LHC and by the future sPHENIX experiment at the RHIC~\cite{Adare:2015kwa}. It is, thus,  important  to pursue  improved  description of  heavy flavor-tagged jet production in heavy-ion collisions using  in-medium higher order 
calculations and resummation. On the experimental side, data in inclusive b-jet production~\cite{Chatrchyan:2013exa} and back-to-back momentum imbalance distributions~\cite{Sirunyan:2018jju} exist. Measurements of c-jets in lead-lead (Pb+Pb) collisions at the LHC are expected in the near future~\cite{hassan:hal-01846896}.

With this in mind, in this paper we will present a calculation of the inclusive charm jet and bottom jet cross sections in heavy-ion collisions using the factorization  Eq.~(\ref{eq:main}) 
based on semi-inclusive fragmenting jet functions for heavy flavor.  We will demonstrate that the final-state  medium-induced  corrections can be incorporated  by modifying the SiJFs,  while the short-distance hard part remains the same as p+p collisions.  These corrections arise from the emergence of in-medium parton showers as the jet 
 evolves in the QCD medium.  We will study these  radiative processes with the help of the medium induced splitting functions, which were calculated up  to the first order in opacity in the framework of SCET with Glauber gluons (SCET$_{\rm G}$)~\cite{Ovanesyan:2011xy} and finite mass effects (SCET$_{\rm M,G}$)~\cite{Kang:2016ofv}. The in-medium splitting functions capture  the full collinear dynamics of energetic parton evolution in a QCD medium. Recent developments based on the formalism of lightcone wavefunctions~\cite{Sievert:2018imd} have allowed us to compute collinear parton branching in QCD media to any order in opacity.  For the purpose of this paper, however,  we will
 restrict ourselves to the first order in opacity results where numerical evaluation of the splitting kernels exists.  Unlike the vacuum case, in the environment of
 strongly interacting matter  jets can dissipate their energy due to collisional interactions with the medium quasi-particles. The collisional energy loss rate can be obtained, for example,
 from the divergence of the energy-momentum tensor of the medium  induced by the color current generated by the jet~\cite{Neufeld:2014yaa}. This work is the first study to include this effect in the SiJF formalism.  Last but not least, we also consider the cold nuclear matter (CNM) and isospin effects~\cite{Kang:2015mta,Chien:2015vja,Albacete:2017qng}.

The rest of our paper is organized as follows:  in Section~\ref{sec:Jet} we discuss the definition of the heavy flavor jet functions in the vacuum. We derive the nuclear modifications of the jet functions in heavy-ion collisions and discuss other relevant nuclear effects.  In Section~\ref{sec:num}  we present our phenomenological results for c-jet and b-jet cross sections in proton and heavy-ion collisions.  Finally, we conclude in Section~\ref{sec:conc}.  

\section{Semi-inclusive jet functions for heavy flavor}
\label{sec:Jet}
In this section, we discuss in detail the definition of the heavy quark-tagged SiJFs. For completeness, we briefly recall some analytical results from Ref.~\cite{Dai:2018ywt}  for the vacuum SiJFs.  We then extend these SiJFs to the case of heavy-ion collisions in the perturbative theory. 

\subsection{Jet Functions in vacuum}
Soft-collinear  effective theory can be generalized to include finite quark masses~\cite{Rothstein:2003wh,Leibovich:2003jd} and is often labeled SCET$_{\rm M}$. Within  SCET$_{\rm M}$,  the heavy flavor SiJFs are obtained by calculating the real and virtual contributions inside the jet cone, and the real contributions outside the jet cone  which are dependent on the jet algorithm. The NLO SiJFs for any parton to produce a heavy flavor-tagged jet,  including the heavy quark mass effects, can be found in Ref.~\cite{Dai:2018ywt}. After renormalization they are given by the following expressions
\begin{align}
    J_{J_Q/Q}(z, p_T R, m, \mu) =& \delta(1-z)+\frac{\alpha_s C_F}{2\pi} \Bigg\{ \delta(1-z)
    \left[ f\left(\frac{m^2}{p_T^2R^2}\right)+g\left( \frac{m^2}{p_T^2R^2}\right)\right]
     \nonumber \\ & 
    - \left(\frac{2z}{1-z}\right)_+ \frac{m^2}{z^2 p_T^2 R^2 + m^2} 
     + \left( \frac{1+z^2}{1-z} \right)_+ \ln\frac{\mu^2}{z^2 p_T^2 R^2 + m^2} 
          \nonumber \\ & \left.
    - \left(2 \frac{1+z^2}{1-z}\ln(1-z)+1-z \right)_+
%     \right.
 %   \nonumber \\ & \left. 
    \right\}~,
    \nonumber \\
    J_{J_s/g}(z,p_T R, m, \mu) = & \delta(1-z) \mathcal{M}^{\rm in-jet}_{g\to Q\bar{Q}}(p_T R, m )+\frac{\alpha_s}{2\pi}\Bigg\{   (z^2+(1-z)^2)
      \nonumber \\ & 
      \left.
    \times 
    \ln \frac{\mu^2}{z^2(1-z^2) p_T^2 R^2 +m^2} 
    -2z(1-z) \frac{z^2 (1-z)^2p_T^2 R^2} {z^2 (1-z)^2p_T^2 R^2+m^2}\right\},  
\end{align}
where the flavor singlet  $J_s = J_Q + J_{\bar{Q}}$.  Here,  $\alpha_s$ is the strong coupling constant, $C_F$ and $C_A$ are the second Casimirs in the
fundamental and adjoint representations of ${\rm SU(3)}$, respectively, $m$ is the heavy quark mass, and $R$ is the jet radius parameter.  The function $J_{J_Q/Q}$ has both  LO and NLO contributions in QCD, while $J_{J_s/g}$ starts at NLO.   The functions $f$ and $g$ have integral representation and are defined in Ref.~\cite{Dai:2018ywt}.  It was argued that, since  heavy quark mass does not affect the ultraviolet (UV) behavior of diagrams, the SiJFs evolve according to DGLAP-like equations similar to the ones for  light SiJFs.  The RG equations
read
\begin{align} \label{eq:dglap}
    \frac{d}{d\ln\mu^2} \left( \begin{array}{cc}
        J_{J_Q/s} (x,\mu) \\
        J_{J_s/g} (x,\mu)
    \end{array} \right)
    =
     \frac{\alpha_s}{2\pi} \int_x^1\frac{dz}{z}  \left(
    \begin{array}{cc}
         P_{qq}(z) & 2 P_{gq(z)}  \\
        P_{qg}(z)  & P_{gg}(z)
    \end{array}
    \right)\left( \begin{array}{cc}
        J_{J_Q/s}  (x/z,\mu) \\
        J_{J_s/g} (x/z,\mu)
    \end{array} \right).
\end{align}
Here, $P_{ij}$ are usual Altarelli-Parisi splitting functions. The evolution equation is solved to leading logarithmic (LL) accuracy using the Mellin moment space approach developed in Ref.~\cite{Vogt:2004ns}. 

In the jet function $J_{J_s/g}(z,m,p_T R, \mu)$ the contribution proportional to $\delta(1-z)$ comes from the $g\to Q\bar{Q}$ splitting inside a jet.  
Starting only at NLO in QCD, it can be written as the integration of the jet fragmentation function $D_{Q/J_g}(z,\mu)$
\footnote{The jet fragmentation function $D_{Q/J_g}(z,\mu,m)$ is defined as the probability to find a Q quark 
with the momentum fraction $z$ inside one given gluon initiated jet. } 
\begin{align}
   \mathcal{M}^{\rm in-jet}_{g\to Q\bar{Q}}(p_T R, m )=&   2 \int_0^1 dz z D_{Q/J_g} (z, p_T R, m). 
\end{align}
As expected, for heavy flavor a new  logarithmic term $\ln p_T R/m$ arises.  When $m \ll p_T R \ll p_T $, in addition to $\ln R$ the logarithmic term $\ln p_T R/m$ needs to be resummed. This can be achieved  through the jet fragmentation function factorization formula  found in Refs.~\cite{Bauer:2013bza,Dai:2016hzf,Dai:2018ywt}. In our case, up to NLO,  $\mathcal{M}^{\rm in-jet}_{g\to Q \bar{Q}}$ can be written as
\begin{align}
    \mathcal{M}^{\rm in-jet}_{g\to Q\bar{Q}}(p_T R, m )=& 2 \sum_{l=g, Q} \bar{K}_{l/g}(p_T R, m, \mu_F ) \bar{D}_{Q/l}(m,\mu_F), 
\end{align}
where $\bar{K}_{l/g}$ is the integrated perturbative kernel at the jet typical scale $p_T R$,  while $ \bar{D}_{Q/l}$ is the integrated parton fragmentation function from parton $l$ to parton $Q$ at the scale of the quark mass. The logarithmic term $\ln p_T R/m$ can be resummed to all order at the LL accuracy by running $D_{Q/l}(m,\mu)$ from $m$ to $p_T R$ analytically.  For further details, we direct the reader to Ref.~\cite{Dai:2018ywt}.

\begin{figure*}
    \centering
    \includegraphics[scale=0.5]{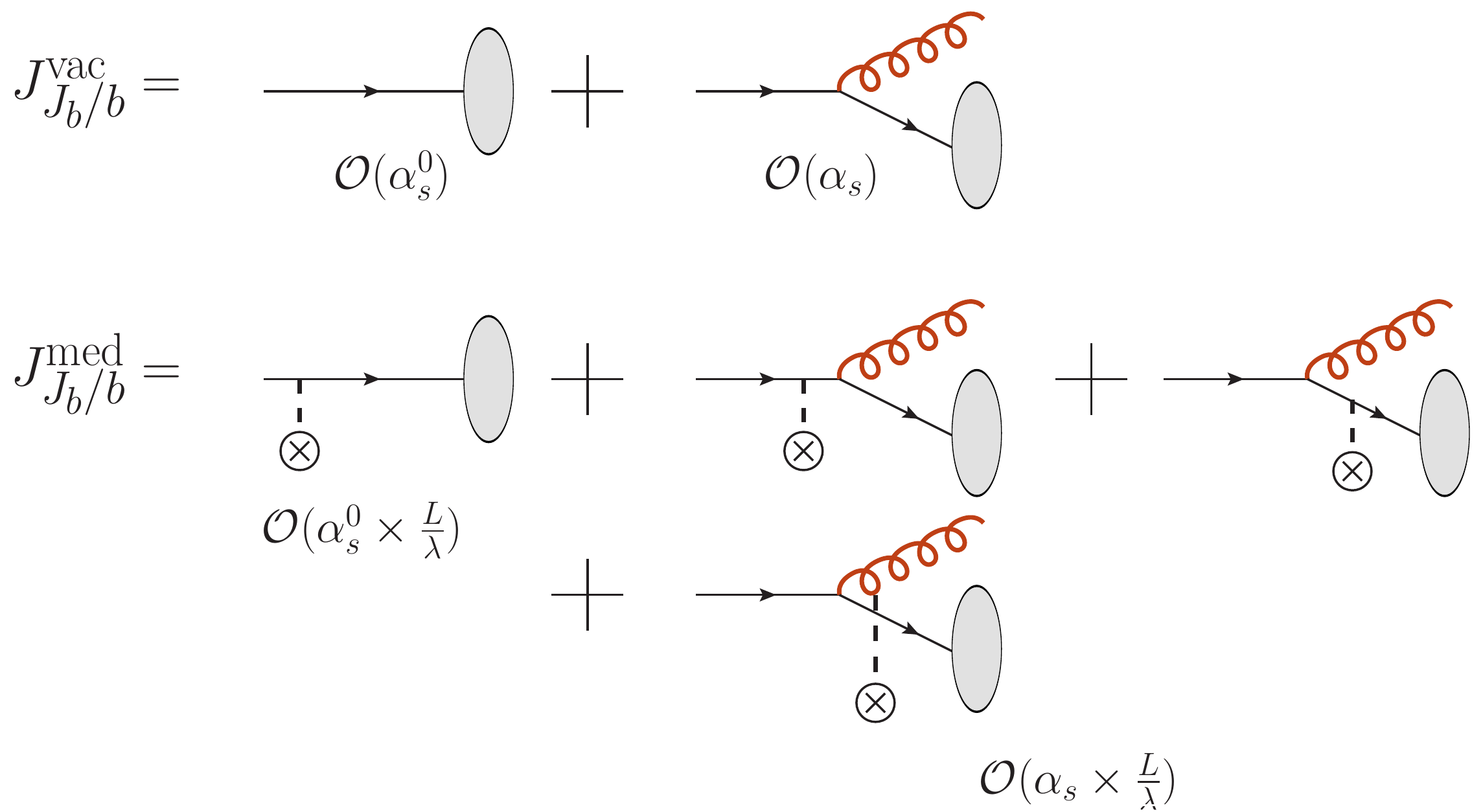}
    \vspace{-0.2cm}
    \caption{Illustration of real diagrams contributing to the $b\to J_{b}$ jet functions in perturbative QCD. The first line shows the diagrams for the LO and NLO contributions to the vacuum jet function, while the second and third lines are the corresponding medium corrections to vacuum diagrams up to the first order in opacity. The dashed lines represent the Glauber gluon exchange between the jet and the QCD medium.  }
    \label{fig:jetfunction}
\end{figure*}

\subsection{Medium corrections }

In perturbative QCD, both of the LO and NLO vacuum SiJFs receive medium modifications, as  illustrated in Fig.~\ref{fig:jetfunction}.    For jets propagating in the forward direction, the interactions with the medium are mediated by $t$-channel Glauber gluon exchanges.  To LO in $\alpha_s$  only in-medium jet  energy dissipation due to collisional  interactions is  allowed. Such collisional energy losses have been evaluated using several different approaches in the literature~\cite{Bjorken:1982tu,Braaten:1991we,Wicks:2005gt,Gossiaux:2008jv,Neufeld:2011yh}. 
At  NLO there is  vacuum radiation along with  corrections from the medium-induced parton shower. 
In the soft gluon approximation, these would correspond to the traditional  radiative energy loss in the QCD medium.  
In principle, one should also consider collisional energy losses along with the medium-induced parton shower.  However, at present there is no theoretical formalism to derive  in-medium splitting functions that simultaneously contain the energy dissipation from the $t$-channel  Glauber gluon exchanges and properly treat the non-Abelian Landau-Pomeranchuk-Migdal effect~\cite{Wang:1994fx}.  The latter is essential to understand the cross section modification of light and heavy particles, and  jets in QCD matter.  Jet physics is particularly challenging, since at higher orders in $\alpha_s$ one needs to identify the specific partons that interact with matter, whether they contribute to the jet function or not, and their precise space-time dynamics.  Branching times would enter in a model-dependent way and, since diagrams with different branching times are combined into  a medium-induced splitting kernel,  even the concept of an average  time  for the splitting to take place is not well defined. It is at this order in $\alpha_s$  
that the model dependence will appear.   Fortunately, both collisional and radiative energy losses are small when compared to the jet transverse energy and their combined effect is suppressed relative to the individual  leading contributions.  We have checked numerically in the limit of long formation times for the medium-induced radiation that the combined effect can give a 10\% correction at the lowest jet $p_T \sim 50$~GeV that we are interested in,  and this correction disappears quickly at high $p_T$. 
For this reason, and because we want to minimize  model dependence in this paper, we include the effect of collisional energy loss in the LO  $\alpha_s^0$  jet function  for  which there is no ambiguity.

We calculate in-medium branching processes to first order in opacity - an approximation that keeps quantum coherence effects between the 
jet production point and any one of the subsequent scattering centers in the QGP. In this framework second order in opacity will include three-body correlations, etc.    
It was found in the soft gluon emission limit that the first order in opacity result is a good approximation for jet quenching calculations at  LHC energies~\cite{Gyulassy:2000fs,Gyulassy:2000er}. This is especially true for realistic plasmas of finite size and density created  in heavy ion collisions and realistic geometries  where a significant fraction of the jets are produced near 
the periphery of the interaction region and traverse thin and dilute media~\footnote{Assumptions of  infinite medium length and dominance of random walk processes might lead to simpler analytic results.  Recent work~\cite{Feal:2018sml}, however,  quantitatively and conclusively showed that an artifact of such assumptions is a gluon emission intensity  that strongly underestimates the high-frequency part of the spectrum and strongly overestimates the low-frequency part of the spectrum. Numerical calculations are, thus, essential to understand in-medium parton branching.}.   It will, of course, be important to study in the future the effects of higher
orders in opacity on the  full medium-induced splitting function using the method  developed in Ref.~\cite{Sievert:2018imd}. These calculations are highly non-trivial, nevertheless we expect second order in opacity splitting function grids to become available in the near future~\cite{Sievert:2019cwq} and to facilitate the evaluation of uncertainties arising  from the opacity expansion.  
In this work we will calculate the medium-corrections to the NLO jet functions up to the first order in opacity which can be written as
\begin{align}
    J^{\rm med}_{J_{Q}/i} =  J^{\rm med, (0)}_{J_{Q}/i} + \frac{\alpha_s}{2\pi} ~J^{\rm med, (1)}_{J_{Q}/i}~.
\end{align}

In this paper we consider for the first time the effects of jet energy dissipation through collisional interactions in the QCD medium in the fragmenting jet function formalism.  
We will limit our results to the LO contribution  $ J^{\rm med, (0)}_{J_{Q}/i}$ and defer the suppressed interplay between collisional and radiative corrections to future work. 
Collisional energy loss effects can be included as a transverse momentum shift of the final state jet which,  to LO in  perturbative QCD, is identified with  the final-state hard parton. The corresponding medium modification to the cross section for the $Q$ jet  production can be written as 
\begin{align}
   \sum_{i} \delta_{iQ} \left[\frac{d\sigma_{pp\to i}(p_T+\delta p_T^i, \eta)}{d(p_T+\delta p_T^i) d\eta } - \frac{d\sigma_{pp\to i }(p_T, \eta)}{d p_T d\eta }  \right]~,
   \label{eq:coll}
\end{align}
where $\delta_{iQ}$ is the Kronecker delta symbol and  $\delta p_T^i$ is the average energy loss for energetic parton $i$ moving through the hot QCD medium. 
As can be seen from Fig.~\ref{fig:jetfunction}, to include only the medium correction here and avoid double counting one should subtract the vacuum contribution. 

We calculate collisional and radiative in-medium effects consistently in a QGP background  simulated  by  2+1-dimensional  viscous  event-by-event  hydrodynamics~\cite{Shen:2014vra}.   For inclusive observables, such as the cross sections of heavy flavor-tagged jets, fluctuations do not play a
role since we average over multiple  hydrodynamic events.  Both collisional and radiative processes in the medium arise from the same $t$-channel interactions between the jet and the medium. The essential information that we use is the  space-time temperature profile of the QGP to evaluate
the transport parameters relevant to jet quenching calculations, see Ref.~\cite{Aronson:2017ymv}.  For example,  $\mu_D^2 = g^2T^2(1+N_f/6)$   is the  Debye screening scale and we use  $N_f = 2$ active light quark flavors. The inverse scattering lengths of  quarks and gluons are 
 ${1}/{\lambda_q}  = \sigma_{qq}\rho_q + \sigma_{qg} \rho_g$, ${1}/{\lambda_g}  = \sigma_{gq}\rho_q + \sigma_{gg} \rho_g$,
where $\rho_q$ and $\rho_g$ are the partial densities of light quarks and gluons in the QGP, respectively.    The elastic scattering cross sections of fast partons  on massive medium quasiparticles are given to lowest order by 
\begin{equation}
\sigma_{qq} = \frac{1}{18 \pi} \frac{g^4}{\mu_D^2} \, ,  \quad \sigma_{qg} = \frac{1}{8 \pi} \frac{g^4}{\mu_D^2} \,   ,  
\quad \sigma_{gg} = \frac{9}{32 \pi} \frac{g^4}{\mu_D^2} \,   .
\end{equation}
In Eq.~(\ref{eq:coll}),  $\delta p_T^i$ is obtained using collisional energy loss rate derived from an operator definition~\cite{Neufeld:2014yaa},  where the QGP constituents are allowed to recoil~\footnote{In Ref.~\cite{Ovanesyan:2011xy} we studied the effect of recoil of quasiparticles  of finite mass on the cross sections  for parton scattering that generates the in-medium branching processes and found this effect to be small, especially at LHC energies.}. We note that 
in heavy ion collisions we work with small radius jets. On the other hand, the dissipated  jet energy is carried away by hydrodynamic medium excitations at angles
${\cal O}(1)$ relative to the direction of jet propagation~\cite{Neufeld:2011yh}. Similar results were obtained in Monte-Carlo simulations of energy flow distributions
inside and outside of jets in Pb+Pb collisions at the LHC~\cite{Tachibana:2017syd}. Thus, we consider this energy fully lost from the point of view of jet production and the jet 
cross section modification can be  equivalently expressed in terms of  the LO  in-medium jet function as
\begin{align}
    J^{\rm med, (0)}_{J_{Q}/i}(z,p_T, \delta p_T^i) = 
    %\\
    z \delta_{iQ} \left[ \delta\left(1-z-\frac{\delta p_T^i}{p_T+\delta p_T^i}\right) -  \delta(1-z) \right]~, 
\end{align}
where only the diagonal part ($i=Q$) is affected by the QGP medium.

The in-medium splitting functions are expressed as  integrals over the medium size and 
the transverse momenta transferred by the Glauber gluon~\cite{Ovanesyan:2011kn,Kang:2016ofv,Sievert:2018imd}. 
Those integrations can only be performed numerically and we use the same QGP background~\cite{Shen:2014vra} 
to obtain grids for all in-medium splitting functions to first order in opacity. 
The medium modification of the SiJFs can  be evaluated from those grids,  
where a UV cut-off $\mu$ rather than the dimensional regularization scheme is used to regularize the UV poles 
when the transverse momentum of the vacuum radiation becomes infinity. 
A similar approach was employed in Ref.~\cite{Kang:2017frl} to construct the SiJFs for massless quarks and gluons. 
The NLO medium correction to the $Q\to J_{Q}$ SiJF is  defined as
\begin{align}
    J^{\text{med},(1)}_{J_Q/Q}(z, p_T R,m, \mu) &=  \int_{z(1-z) p_T R }^{\mu} d q_\perp P^{\rm med}_{QQ}(z, m, q_\perp) 
    \nonumber \\ & % \hspace{-3cm}
     - \delta(1-z) \int_0^1 dx  \int_{x(1-x)  p_T R}^{\mu} d q_\perp P^{\rm med}_{QQ}(x, m, q_\perp)   
    \nonumber \\ &
    =    \left[\int_{z(1-z)  p_T R}^{\mu} d q_\perp P^{\rm med}_{QQ}(z, m, q_\perp) \right]_+~.
\end{align}
In the above equation, the first line corresponds to the contribution with a radiation outside of the jet cone, while the second line represents the combination of the real radiation inside the jet cone and the virtual loop corrections.   The singularity that arises  when $z\to 1$ is regularized properly by the plus distribution function after combining all the corrections. 
The medium correction to the channel $g\to J_s = J_{Q}+J_{\bar{Q}}$ is 
\begin{align}
    J^{\text{med},(1)}_{J_s/g}(z, p_T R, m, \mu) = & 
 %   \nonumber \\ &  \hspace{-3.5cm}
     \int^{\mu}_{z(1-z) p_T R} dq_\perp \left(P^{\rm med}_{Qg}(z,m, q_\perp)+  P^{\rm med}_{\bar{Q}g}(z,m, q_\perp)  \right)
        \nonumber \\ &
    +\delta(1-z) \int_0^1 dx \int_0^{x(1-x) p_T R} dq_\perp  P^{\rm med}_{Qg}(x,m, q_\perp)~,
\end{align}
where the first and second lines are the contributions with the real radiation outside and inside the jet cone, respectively.

In the application of in-medium splitting functions we use extensively  sum rules. 
Let us take for example the momentum sum rule for the gluon initiated splitting 
\begin{align}
  \int_0^1 x~dx \left( P_{gg}(x) + \sum_{i} P_{ig}(x)   \right )  = 0~.
 \end{align}
In this formulation $\sum_i$ runs over all quark and antiquark flavors, therefor we don't have an explicit $2n_f$ factor. 
Since the sum rules are satisfied by the vacuum part separately,  
they are also satisfied by the medium-induced splitting kernels.  
 Up to the NLO,  in our previous works~\cite{Chien:2015vja,Kang:2016mcy}  
 we have implemented  the following normalization for the splitting functions 
\begin{align}
   \int_0^\mu dq_\perp \int_0^1 x~dx \left( P^{\rm med}_{gg}(x) + \sum_{i} P^{\rm med}_{ig}(x)   \right)  = 0~,
 \end{align}
 where the $ \sum_{i}$ runs over all the light flavor quarks and the $q_\perp$ is the transverse momentum of the radiation.  In order to keep the sum rule and consistency with 
 our earlier calculations,  for the heavy flavor  splitting function we should have
\begin{align} \label{eq:pqg}
  \int_0^\mu dq_\perp \int_0^1 x ~dx P^{\rm med}_{Qg}(x, q_\perp) = 0 
\end{align}
Using the symmetry $P_{Qg}(x, q_\perp) =  P_{Qg}(1-x, q_\perp)$ we have
\begin{align}
  0  =  &  \; \; 2    \int_0^\mu dq_\perp \int_0^1 x~ dx P^{\rm med}_{Qg}(x, q_\perp)     \nonumber \\  \hspace{-4cm}
     =   &  \; \;  \int_0^\mu dq_\perp \int_0^1 x ~dx P^{\rm med}_{Qg}(x, q_\perp) 
%      \nonumber \\  & \;\; 
      +  \int_0^\mu dq_\perp \int_0^1 (1-x) ~dx P^{\rm med}_{Qg}(x, q_\perp)      
          \nonumber \\  \hspace{-4cm}
       =  & \; \;    \int_0^\mu dq_\perp dx  P^{\rm med}_{Qg}(x, q_\perp)~.
\end{align}
The equation above implies that there is no additional production of open heavy flavor per nucleon-nucleon collision in heavy ion reactions. This is 
consistent with experimental measurements~\cite{Adcox:2002cg}.  
Taking this constraint into account,  the function $J^{\text{med},(1)}_{J_s/g}$ becomes
%\begin{widetext}
\begin{align}
    J^{\text{med}, (1)}_{J_s/g}(z, p_T R, m, \mu) 
    = & \left[ \int_{z(1-z)  p_T R}^{\mu} d q_\perp P^{\rm med}_{Qg}(z, m, q_\perp) \right]_+
    \nonumber \\ & 
    +   \int^{\mu}_{z(1-z) p_T R} dq_\perp P^{\rm med}_{Qg}(z,m, q_\perp)~.
\end{align}
%\end{widetext}

The full in-medium SiJFs are defined as 
\begin{align}
    J_{J_Q/i} = J_{J_Q/i}^{\rm vac, LL} +  J_{J_Q/i}^{\rm med} \, , 
\end{align}
where the vacuum contributions are calculated at the LL  accuracy, while only the fixed-order medium corrections are included consistently. In this case we choose the cut-off scale $\mu$ as the jet's transverse momentum.  In principle the full in-medium SiJFs obey a DGLAP-like evolution in Eq.~(\ref{eq:dglap}) and  one can even introduce the medium-induced splitting functions in the DGLAP kernel to fully consider the jet evolution in the QCD medium.  We will leave this for a future study. 

Besides final-state in-medium modifications to jet production, we have to include the initial-state ones,  or CNM  effects. The CNM  effects arise from the coherent, elastic,  and  inelastic patron scattering in large nuclei  only part of them can be parameterized via global nuclear  PDF fits,  for example nCTEQ15 PDF sets~\cite{Kovarik:2015cma}.   While many of these effects  can be implemented  as {\em effective} modifications of the PDFs, they are  not necessarily universal.  A well known example is the Cronin effect which manifests itself in the enhancement of
the cross sections for particle production in proton-nucleus  (p+A) reactions relative to p+p reactions at transverse momenta in the range  $p_T = 1 - 5$~GeV~\cite{Cronin:1974zm}.  As this enhancement range is independent of the center of mass energy  of the collision, this effect does not scale with the momentum fraction of the partons in the nucleus. The most common theoretical explanation of the Cronin effect is related to the initial-state transverse momentum broadening of the partons  through $t$-channel interactions  with cold nuclear matter~\cite{Zhang:2001ce,Accardi:2001ih, Vitev:2002pf}.  As we have seen in the case of the QGP, these interactions will lead to medium-induced radiation. Its spectrum was  first computed for beam jets by Bertsch and Gunion~\cite{Gunion:1981qs} to provide a perturbative explanation of the nearly flat  rapidity distribution of soft particles produced in hadronic collisions. Naturally, in p+A collisions the radiation intensity will be amplified  and, as hard partons lose a small fraction of their energy, the cross sections for particle and jet production can be reduced. The first measurement to support this energy loss picture was performed at a Fermilab  fixed target experiment~\cite{Leitch:1999ea} for $J/\psi$s, where the strong suppression of the cross section was  found to  scale with the momentum fraction of the parton from the proton and not with the momentum fraction of the parton from the nucleus. This shows conclusively that the effect does not arise from nuclear PDF modification, 
but is in line with the expectations from the picture of parton energy loss in large nuclei~\cite{Gavin:1991qk,Vogt:1999dw}.  The same scaling and physical picture were later confirmed at collider energies~\cite{Adler:2005ph,Kopeliovich:2005ym}.   
  
Most recently, measurements of jets at RIHC and LHC in deuterium-gold (d+Au) and proton-lead (p+Pb), respectively,  have also revealed cross suppression in central and semi-central collisions~\cite{ATLAS:2014cpa,Adare:2015gla}. This experimentally  determined modification of the jet cross sections was also found to scale with a variable proportional to the momentum 
fraction of the parton in the proton and evaluated in a picture of parton energy loss in cold nuclear matter~\cite{Kang:2015mta}.  The radiation from the beam jets was numerically calculated as in
Ref.~\cite{Vitev:2007ve} and is implemented as a shifts in Bjorken $x$ variable  in the PDFs as follows:
\begin{align}
    f_{q/A}(x,\mu) &\to f_{q/A}\left(\frac{x}{1-\epsilon_q},\mu\right) , 
    \qquad
    f_{g/A}(x,\mu) \to f_{g/A}\left(\frac{x}{1-\epsilon_g},\mu\right)~.
\end{align}
This shift is related to the energy fraction $\epsilon_{q,g}$  that the parton loses in cold nuclear matter and depends on the parton flavor. The cross section suppression effect will be the
largest for large values of Bjorken $x$ because of the steeply falling parton distributions as a function of  $x$. One example that we have studied besides jets is the attenuation of di-muons 
in fixed target Drell-Yan production~\cite{Neufeld:2010dz}. For the smaller jet $p_T$ ranges that we consider  here relative to the ATLAS p+Pb measurement~\cite{ATLAS:2014cpa}  the effect will be reduced, as we move 
away from the kinematic bounds. However,  in nucleus-nucleus (A+A) collisions it will be amplified since partons form both nuclei pass through the opposite going nucleus.  
In this calculation we also have  control over the centrality dependence of the CNM effects, for example they are $\sim20\%$ bigger per nucleus in central relative to 
minimum bias  Pb+Pb collisions.  
For the case of nuclei, we include with relevant weights $Z$ and $A-Z$ the  proton  and neutron PDFs,  respectively,  where the latter are constructed from the proton PDFs using isospin symmetry.  
The validation of our evaluation of CNM effects will be discussed in Fig.~\ref{fig:RpA} by comparing  the predicted $R_{\rm pA}$ to CMS measurements.

\section{Numerical results } \label{sec:num}

In this section we present our numerical predictions for inclusive c-jet and b-jet production in hadronic collisions. For our work we choose the CT14NLO PDF sets~\cite{Dulat:2015mca}. The hard part in the factorized cross section is calculated at NLO with massless b- and c-quarks,  while the mass effect is included in the c- and b-jet SiJFs.  The UV cut-off and the scale of  $\alpha_s$ in the medium corrections to the jet functions are set to be the factorization scale $\mu$.  The default factorization and jet scales are chosen to be  $\mu=p_{T}$ and $\mu_J= p_T  R$, respectively. The uncertainties are evaluated by varying $\mu$ and $\mu_J$ by a factor of 2 independently. The coupling between the jet and the medium, which appears in the modification to the SiJFs, is set as $g=2$~\footnote{ The average strong coupling constant is $\bar{\alpha}_s=g^2/4\pi=0.32$ consistent with many other studies on jet quenching effects~\cite{Huang:2013vaa,Chien:2015vja,Chien:2015hda,Kang:2017frl,Li:2017wwc,Kang:2018wrs}. Comparison between predictions based on the medium-induced splitting functions and data might provide a way to extract the coupling $\bar{\alpha}_s$, though one should be mindful of theoretical model uncertainties.}.

\begin{figure*}[!t]
    \centering
    \includegraphics[width=0.49\textwidth]{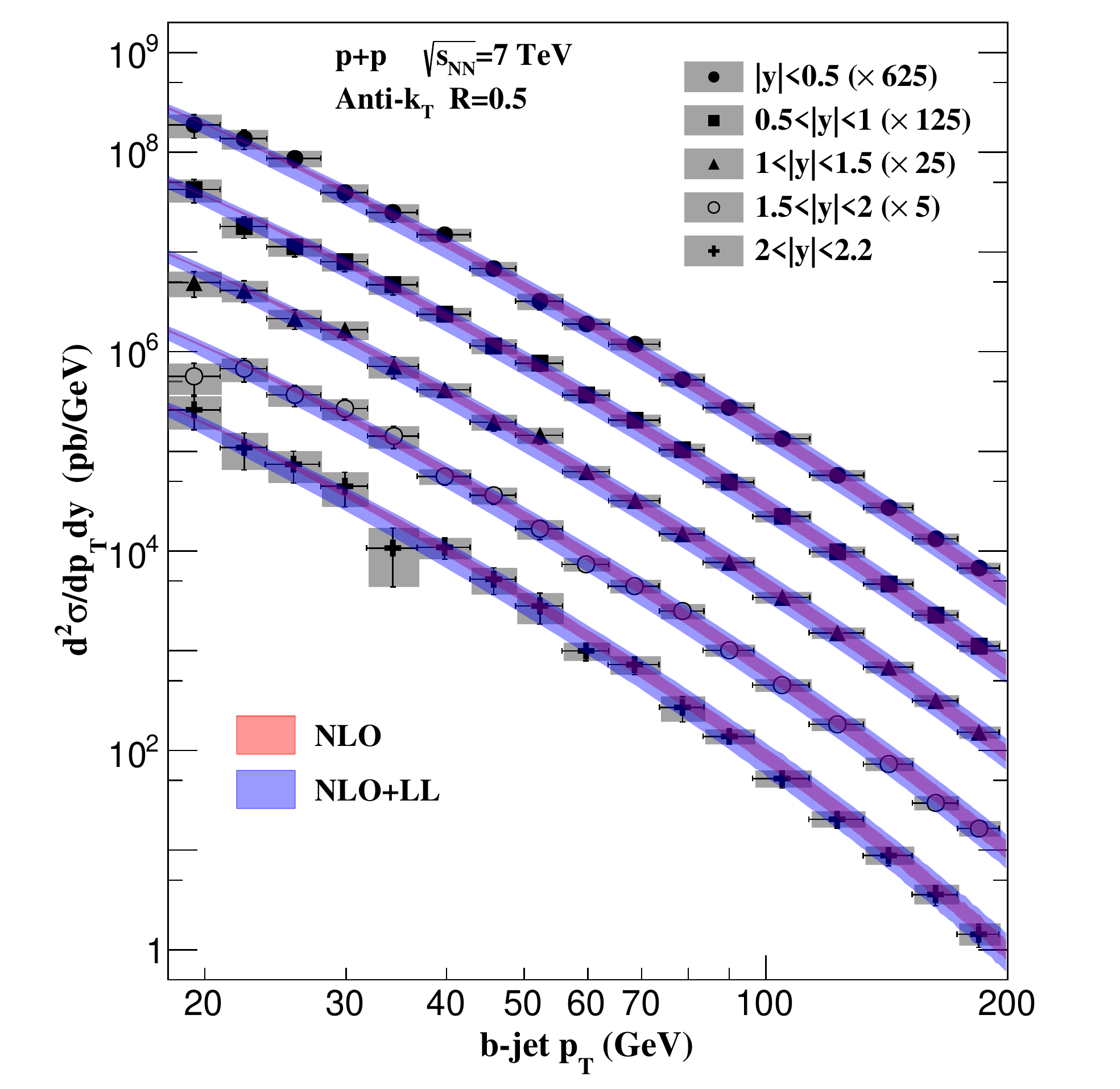}
    \includegraphics[width=0.49\textwidth]{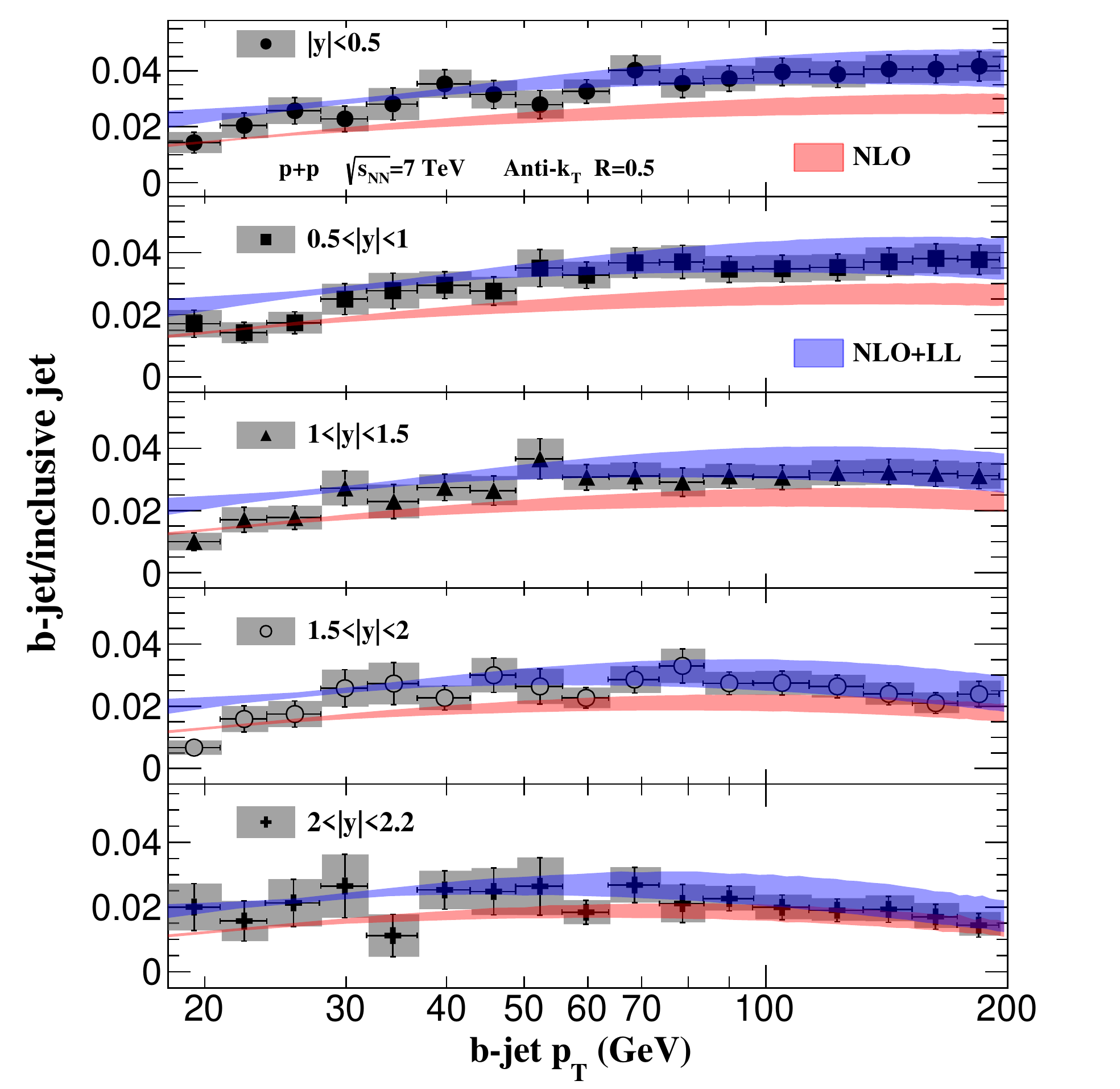}
    \vspace{-0.5cm}    
    \caption{ Comparison of theoretical predictions and  measurements for b-jets with radius $R=0.5$ in p+p collisions at $\sqrt{s_{\rm NN}}=7$ TeV. The b-jet cross section (left) and the ratio of the b-jet cross section to the inclusive jet cross section (right) are shown as a function of jet $p_T$ for different rapidity intervals.  The red and blue bands are the NLO and NLO+LL predictions, respectively, while the gray bars represent the total experimental uncertainty of the data~\cite{Chatrchyan:2012dk}. 
     }
    \label{fig:pp_b_cms}
\end{figure*}

Before we move  on to  jet production in heavy-ion collisions, we must address inclusive b-jet and c-jet production in p+p collisions. These cross sections set the baseline relative to which cross section modifications with nuclei can be detected.  Figure~\ref{fig:pp_b_cms} presents the comparison between our theoretical results for the inclusive b-jet cross section (left) and the fraction of b-jets to inclusive light jets  (right) for different rapidity intervals as a function of the jet transverse momentum $p_T$.  The colliding system is p+p with $\sqrt{s_{\rm NN}}$=7~TeV and jet reconstruction parameter R=0.5. 
The $\ln R$ resummed cross sections for b-jet changes the NLO predictions by ${\cal O}(10\%)$, but we find larger scale uncertainties. The NLO and NLO+LL $p_T$ distributions presented here are consistent with the predictions from Ref.~\cite{Dai:2018ywt} and the experimental measurements~\cite{Chatrchyan:2012dk}. For the b-jet fraction, the difference between NLO+LL and NLO visible in the right panel of  Figure~\ref{fig:pp_b_cms} can be traced also to the differences in the inclusive jet cross section. The $\ln R$ resummation reduces the inclusive jet cross more significantly.  The NLO+LL predictions agree very well with the data.

\begin{figure*}[!t]
    \centering
    \includegraphics[width=0.49\textwidth]{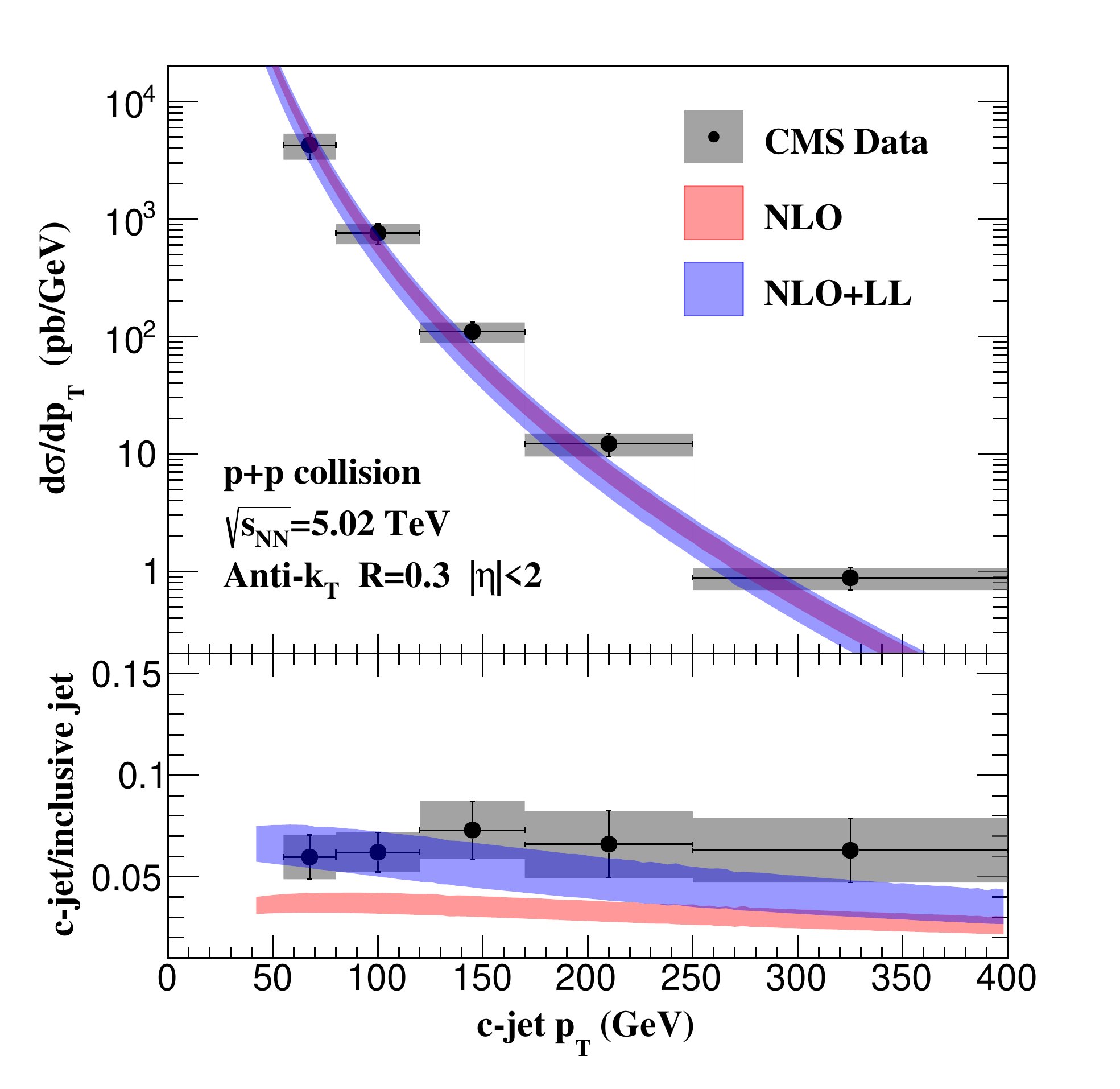}
    \includegraphics[width=0.49\textwidth]{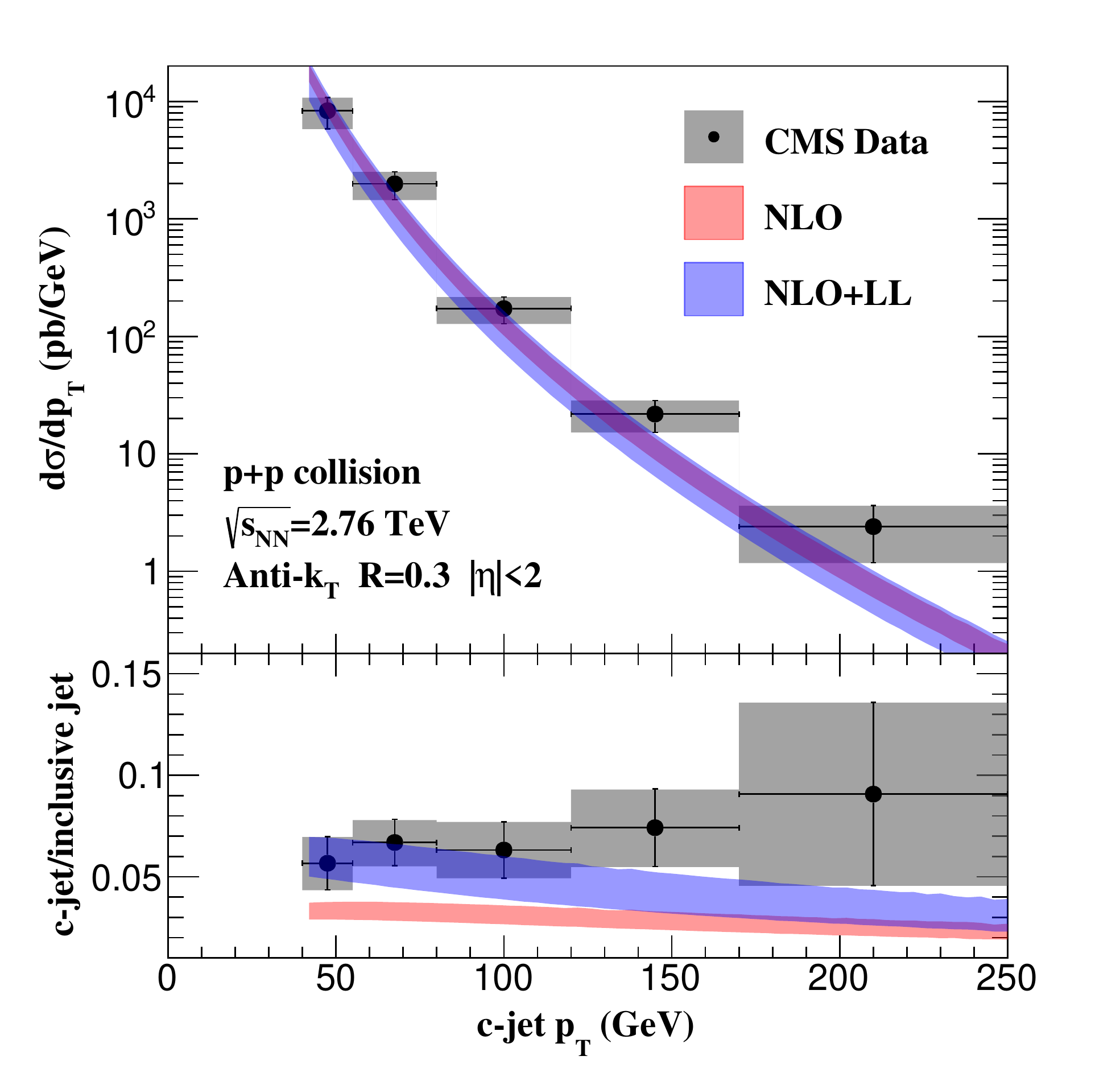}
    \vspace{-0.5cm}
    \caption{The c-jet cross sections (top) and their ratios (bottom) to the inclusive jet cross sections as a function  of jet $p_T$ in p+p collisions at 
    $\sqrt{s_{\rm NN }}=5.02$~TeV (left) and $2.76$~TeV (right).  The jets are reconstructed using the anti-k$_{\rm T}$ algorithm with R=0.3 and the pseudorapidity cut $|\eta|<2$ is imposed. 
    The color scheme is the same as in Fig.~\ref{fig:pp_b_cms}.  }
    \label{fig:pp_c_cms}
\end{figure*}

We now move on to c-jets and their cross sections in p+p collisions are shown in Fig.~\ref{fig:pp_c_cms} for center-of-mass energies 5.02~TeV (left) and 
2.76~TeV (right), respectively.  Similar to b-jet cross section, the $\ln R$ resummation changes the c-jet cross section by about 10\% in a $p_T$-dependent fashion,   but with larger theoretical uncertainties. We also note that in comparison to b-jets the spectrum appears stiffer, and our calculation agrees better with the experimental measurements at lower $p_T$.  This is also reflected in the right panel  of Fig.~\ref{fig:pp_c_cms}.    Again, the NLO+LL c-jet fractions agree better with CMS measurements~\cite{Sirunyan:2016fcs} when compared to the NLO ones for both collision energies. 

The high energy jet production in proton-nucleus collisions can place constraints on the  CNM effects. Early ATLAS and PHENIX measurements suggested that 
the suppression of inclusive jet cross sections, especially at high $p_T$ and in central p+Pb collisions~\footnote{Determination of centrality in p+A collisions can be subtle, thereby influencing the experimentally measured nuclear modification.},  can be large~\cite{ATLAS:2014cpa,Adare:2015gla}.  The ALICE collaboration as further studied the event activity in semi-inclusive  hadron-jet distributions~\cite{Acharya:2017okq}. In minimum bias collisions, the modification of jet cross sections, if any, is smaller. Within the statistical and systematic uncertainties many measurements are consistent with a range of possibilities - from no nuclear effects to $\pm10\%$ cross 
section modification. The way in which the modification due to in-medium QCD effects is experimentally and theoretically studied is through the  ratio
\begin{equation}
R_{\rm AB} = \frac{1}{\langle  N_{\rm bin}^{\rm AB} \rangle} \frac{ d\sigma^{\rm AB}/d({\rm PS})}{d\sigma^{pp}/d({\rm PS})}\; .
\label{RAB}
\end{equation}
Here, $\langle  N_{\rm bin}^{\rm AB} \rangle $ is the average number of binary collisions between proton/nucleus A and proton/nucleus B and
$d({\rm PS})$ is the relevant phase space differential - in our case in the jet transverse momentum $p_T$. An added advantage of the observable 
defined in Eq.~(\ref{RAB}) is that uncertainties in the absolute normalization of the cross section largely cancel in the ratio. 

\begin{figure}
    \centering
    \includegraphics[width=0.5\textwidth]{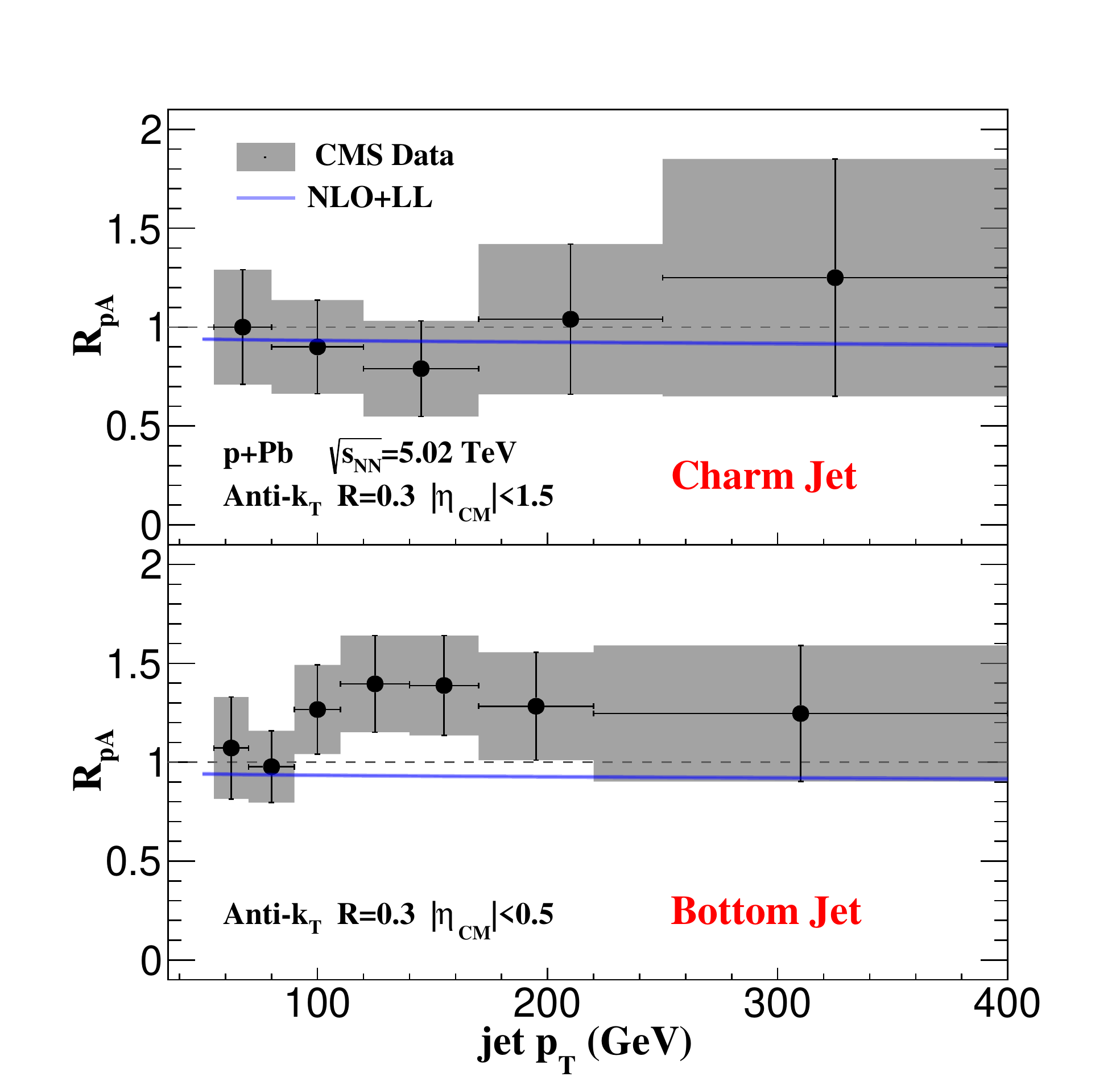}
    \vspace{-0.5cm}
    \caption{Comparison of theoretical predictions for the heavy flavor jet cross sections in proton-nucleus collisions $R_{\rm pA}$ to CMS measurements~\cite{Sirunyan:2016fcs,Khachatryan:2015sva} of c-jets (top) and b-jets (bottom). The system is p+Pb at $\sqrt{s_{\rm NN}}=5.02$~TeV and the jets are constructed using the anti-k$_{\rm T}$ algorithm with $R=0.3$. The rapidity cut in the center-of-mass frame is $|\eta_{\rm CM}|<1.5$ for c-jets and $|\eta_{\rm CM}|<0.5$ for b-jets. }
    \label{fig:RpA}
\end{figure}

We  can use proton-lead collisions at the LHC  to check the validity of our theoretical model. At the high jet transverse momenta under investigation, we take into  
account the initial-state CNM energy loss~\cite{Kang:2015mta}.
In Fig.~\ref{fig:RpA}, we compare our calculated nuclear modification factor $R_{\rm p A}$ for heavy flavor tagged-jets to the CMS experimental measurements~\cite{Sirunyan:2016fcs,Khachatryan:2015sva} in 5.02~TeV p+Pb collisions. The scale dependence of the cross sections in p+Pb collisions is almost the same as the one in p+p collisions. 
Therefore, the NLO+LL  ratio $R_{\rm p A}$ changes very little with scale variation. Furthermore there is not an obvious difference between the predicted c-jet and b-jet $R_{\rm pA}$, which is between 0.9 and 0.95 for the jet transverse momentum in the range of 50 GeV $<p_{T}<$ 400 GeV. The NLO+LL jet cross section modification  $R_{\rm pA}$  for c-jet, displayed in the top panel,  describes the CMS data~\cite{Sirunyan:2016fcs} very well. In the bottom panel, the prediction is on the lower edge of the experimental error bar~\cite{Khachatryan:2015sva} and cannot describe the fluctuation of the data around $p_T\sim150$ GeV. Given  experimental uncertainties, however, there is no significant deviation between predictions and the measurements.  It is worth noticing that in Pb+Pb collisions  CNM effects will be amplified when compared to p+Pb collisions because there is one more nucleus in the initial state.  

\begin{figure*}
    \centering
    \includegraphics[width=0.7\textwidth]{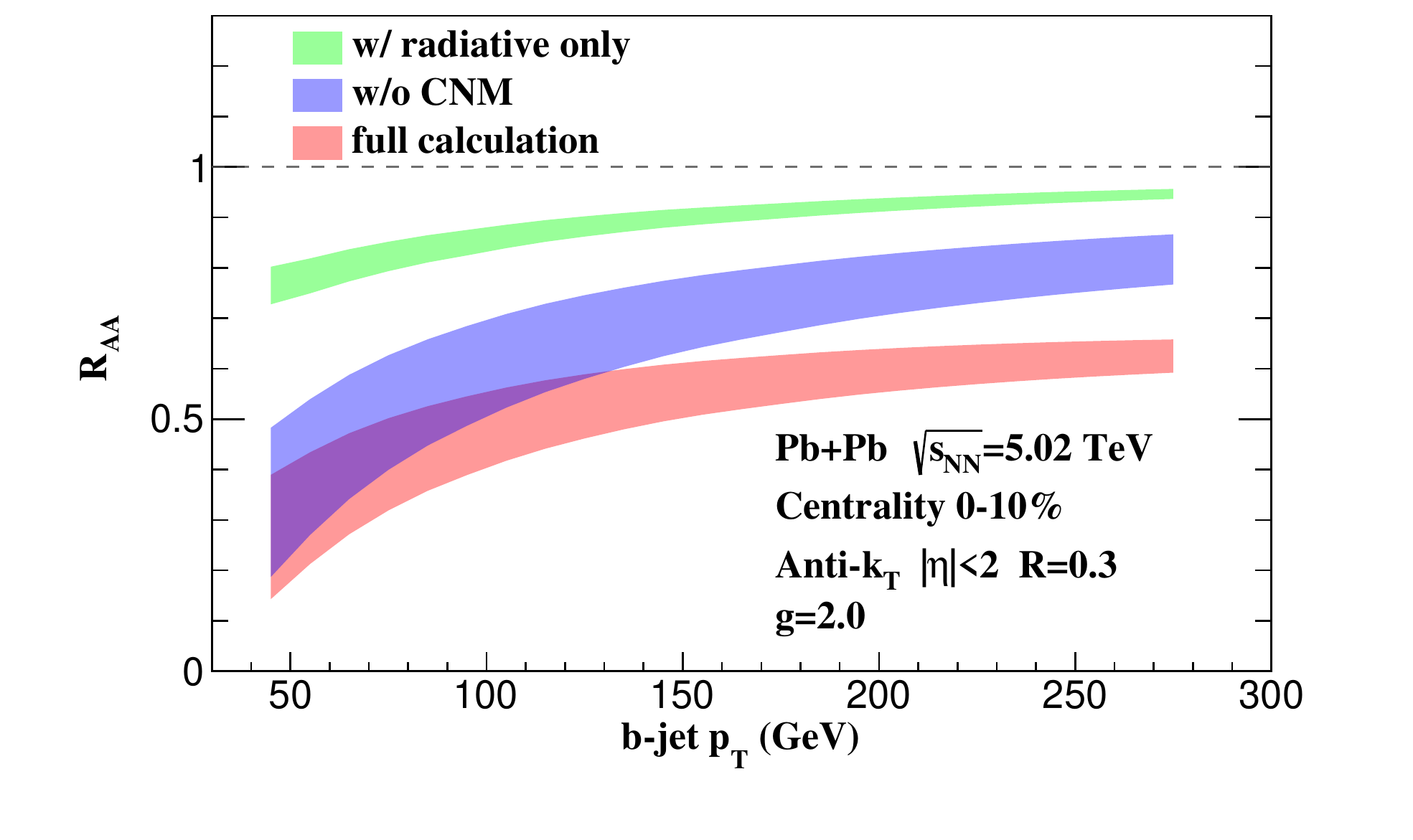}
    \vspace{-0.8cm}
    \caption{The b-jet suppression factor for  central Pb+Pb collision at $\sqrt{s_{\rm NN}}$=5.02 TeV, where anti-k$_{\rm T}$ algorithm with R=0.3 and jet $|\eta|<2$ are implemented.  The $R_{\rm AA }$ with all nuclear effects included is shown as the red band. The blue band presents the prediction without the CNM effects, while the green band corresponds to the result with only  in-medium  radiative  processes included.    }
    \label{fig:raa_b_5}
\end{figure*}

To proceed to nucleus-nucleus reactions, we include final-state interactions.
In Fig.~\ref{fig:raa_b_5} we show the theoretical model predictions for the in-medium suppression factor $R_{\rm AA}$ for  b-jets with $|\eta|<2$,  reconstructed with 
the anti-k$_{\rm T}$ algorithm with R=0.3  at $\sqrt{s_{\rm NN}}$=5.02 TeV.  The effect of the medium-induced parton shower is represent by the green band. 
Compared to the light jet results from Ref.~\cite{Kang:2017frl},  the effect of in-medium radiative processes on b-jets is noticeably smaller. The reason for that lies in the strength  of the medium-induced parton shower contribution to b-jet production,  which is predominantly proportional to the second Casimir in the color representation of the parent
parton and is smaller for quark-initiated jets.  The difference between the blue and green bands in Fig.~\ref{fig:raa_b_5}  represents the jet energy dissipation in the medium
due to  collisional  processes.  It is of the same order as the medium-induced out-of cone radiation  and is more important when the jet transverse momentum is small. 
The CNM effects, represented by the  difference between the blue and red bands, are more important in the high energy regime, especially  when the attenuation due to final-state
effects become smaller.  As expected,  they are more than twice larger than the $R_{\rm pA}$ in Fig.~\ref{fig:RpA} since we consider central Pb+Pb collisions. The full nuclear modification factor $R_{\rm AA}$ is about 0.3 for $p_T\sim$ 50 GeV and about 0.6 for $p_T \sim$ 250 GeV. Even though the b-jet $R_{\rm AA}$ is found to be qualitatively consistent with that of inclusive jets from measurements with the same collision energy~\cite{Aad:2012vca,Khachatryan:2016jfl}, the underlying physics might be different.

\begin{figure*}
    \centering
    \includegraphics[width=0.49\textwidth]{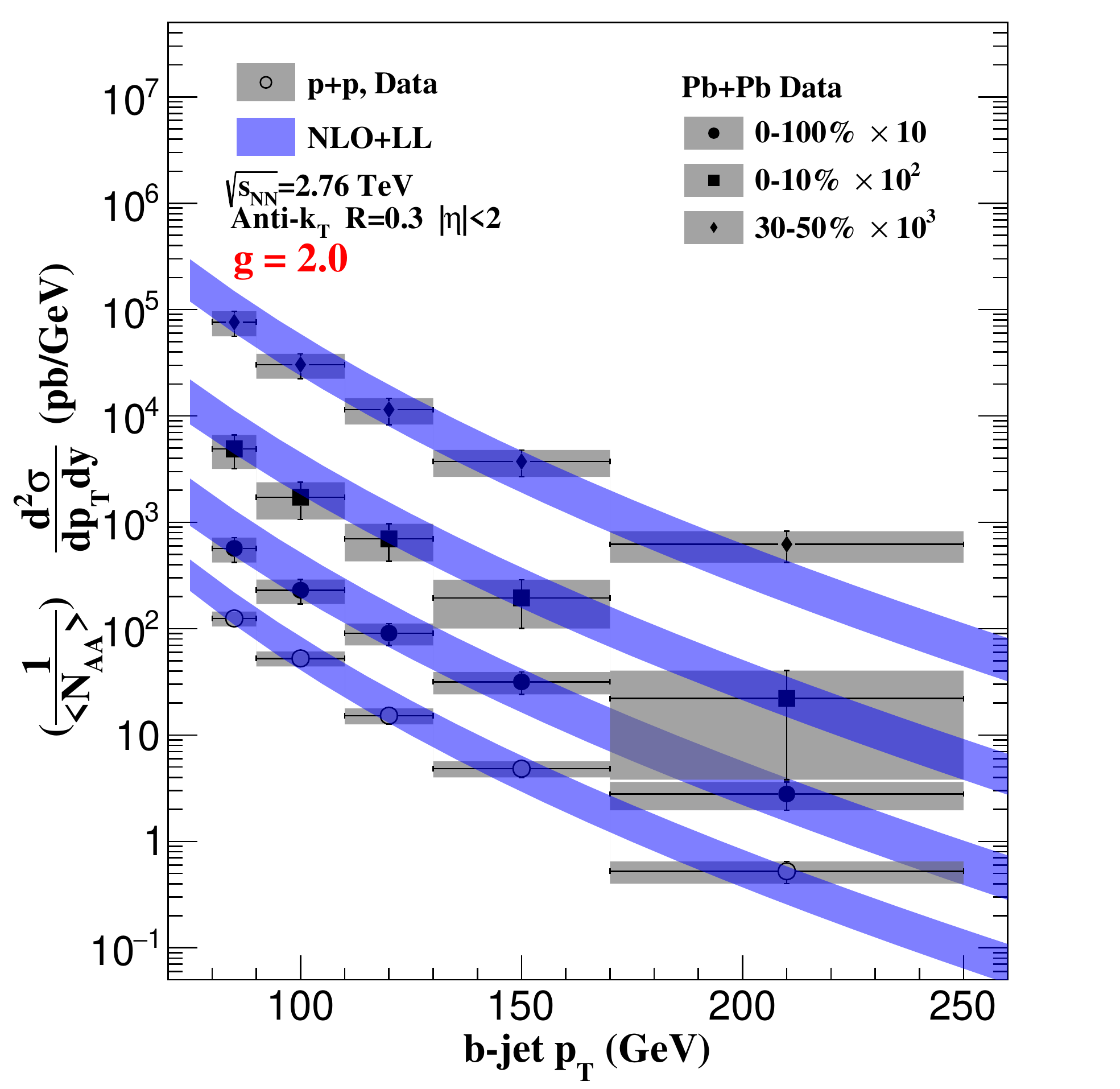}
    \includegraphics[width=0.49\textwidth]{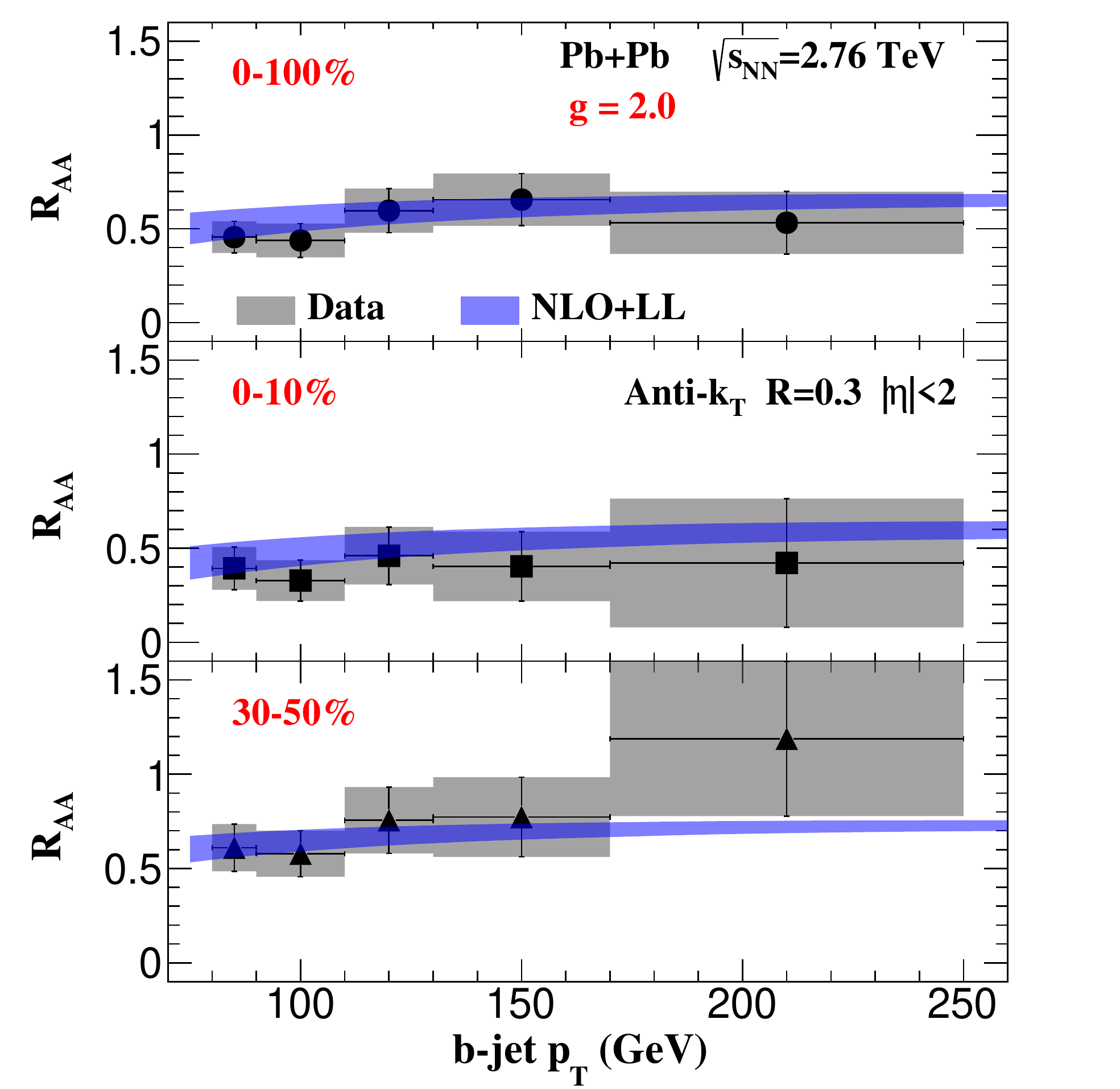}
        \vspace{-0.1cm}
    \caption{The b-jet cross sections (left) in p+p and Pb+Pb collisions and the nuclear modification factor $R_{AA}$ (right) for different centrality classes (0-100\%, 0-10\% and 30-50\% ), as indicated  in the legend. All theoretical predictions are present as the blue bands and are compared to the data from CMS measurements~\cite{Chatrchyan:2013exa}.  }
    \label{fig:pbpb_b_cms}
\end{figure*}

\begin{figure*}
    \centering
    \includegraphics[width=0.49\textwidth]{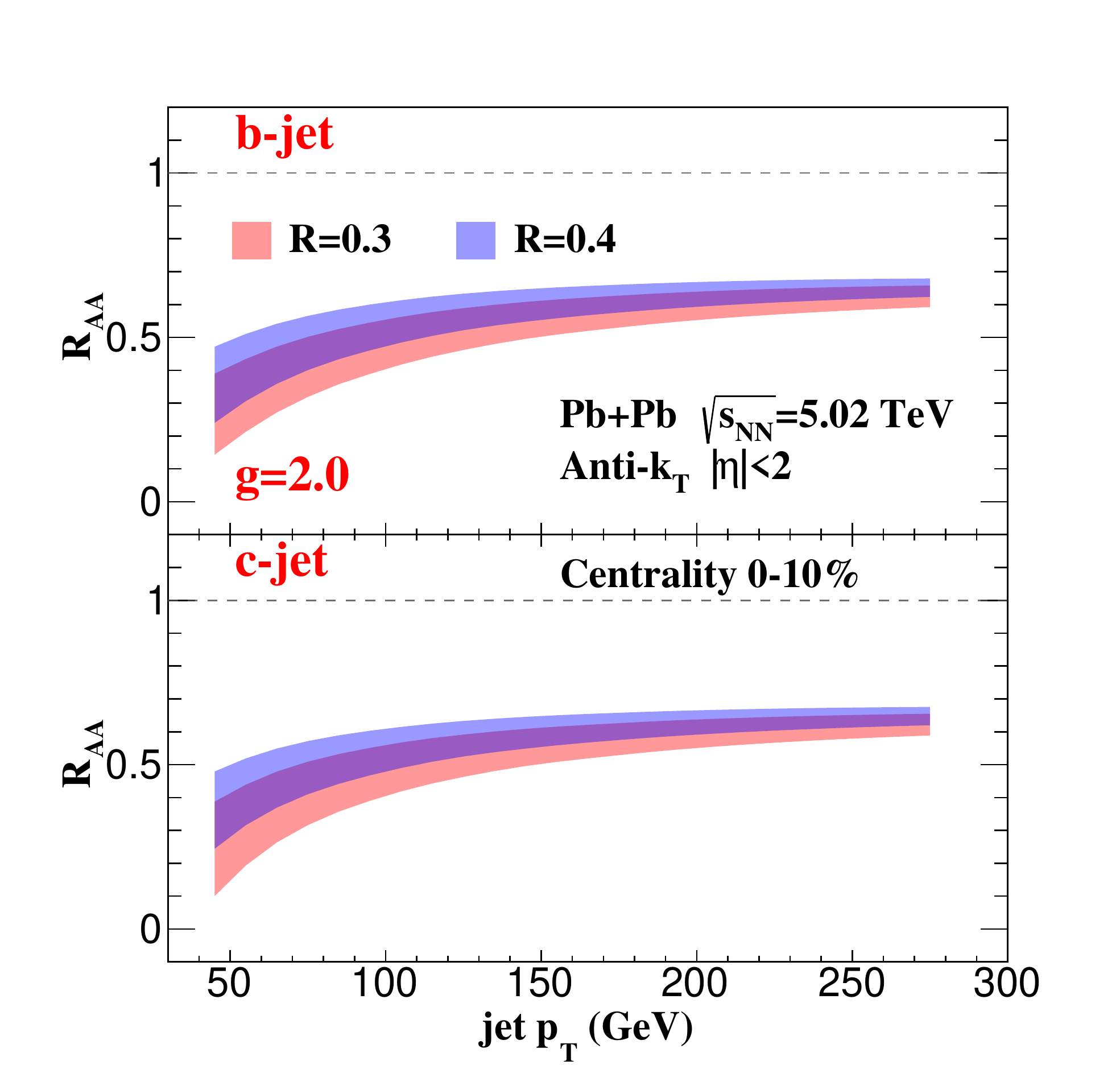}
    \includegraphics[width=0.49\textwidth]{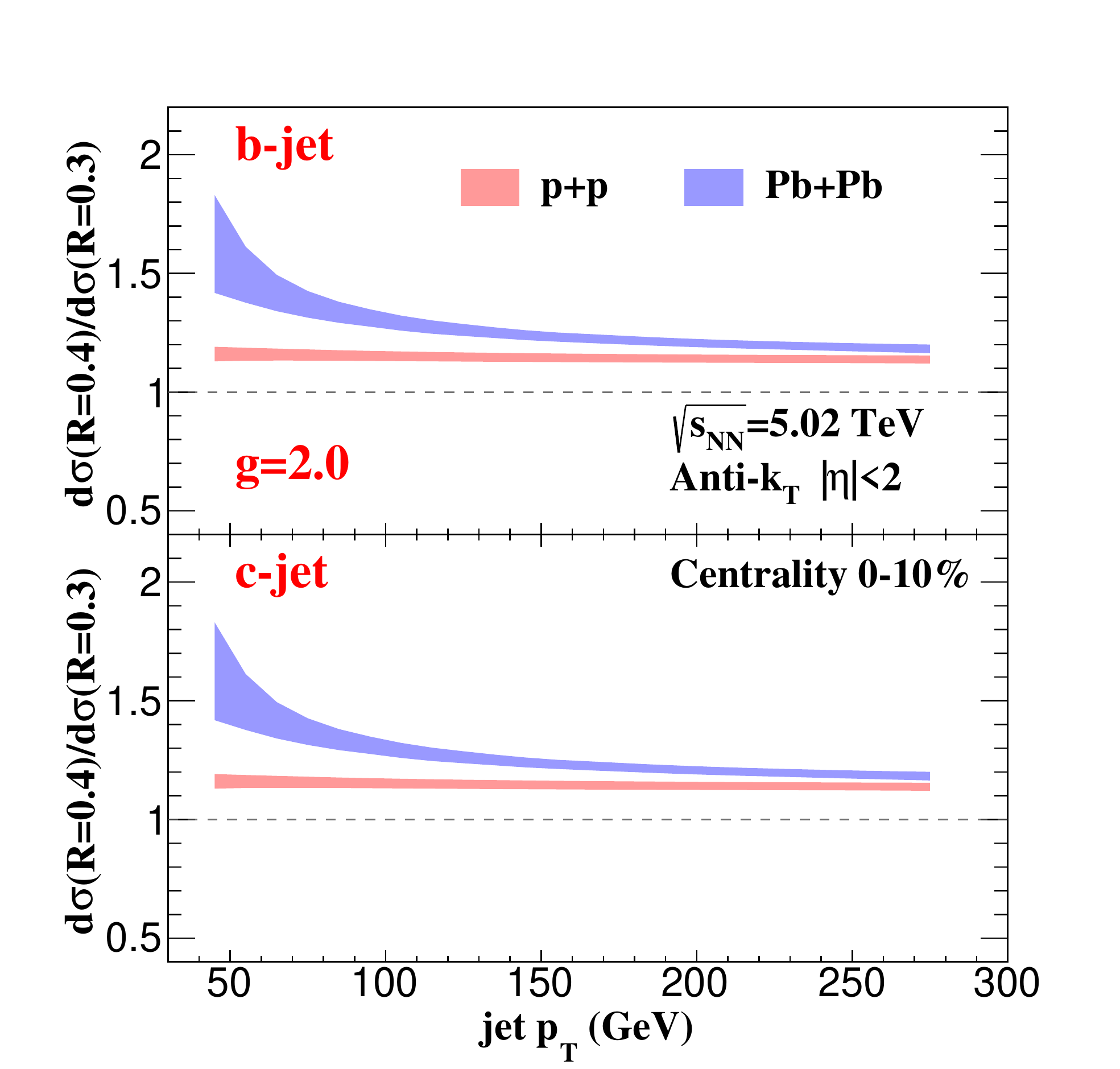}
    \vspace{-0.1cm}
    \caption{The  heavy flavor tagged jet  nuclear suppression factor $R_{\rm AA}$ (left) and the cross section ratio for different radii (right).   In the left plot, the red and blue bands represent the $R_{\rm AA}$ for R=0.3 and 0.4, respectively. The  cross section ratio is defined as $\frac{d\sigma(R=0.4)}{dp_T}/\frac{d\sigma(R=0.3)}{dp_T}$ with the red and blue bands for p+p and Pb+Pb collisions, respectively.
    The upper panels present our results for b-jet and the lower ones for c-jet. 
    }
    \label{fig:Raa5}
\end{figure*}

In Fig.~\ref{fig:pbpb_b_cms} we compare our numerical calculations to the measurements~\cite{Chatrchyan:2013exa} of b-jet production in p+p and Pb+Pb collisions with $\sqrt{s_{\rm NN}}=2.76$ TeV.  The left plot shows the inclusive b-jet $p_T$ distributions, where the cross sections in heavy-ion collisions with centrality 0-100\%, 0-10\% and 30-50\% are scaled by powers of 10 for visibility. The right plot presents the nuclear modification factor $R_{\rm AA}$ with the same settings. In general $R_{\rm AA}$ decreases (larger suppression) with increasing collision centrality (toward head-on  nuclear collisions). From the data, the attenuation factor $R_{\rm AA}$ seems less dependent on the centrality when compared to the well-known light jet modification. The predictions agree very well with the data for both  the inclusive cross sections and the nuclear modification factors.
QCD medium corrections depend on the center-of-mass energy,  colliding nuclei, and the  centrality of the collision 
itself. Thus, the agreement between data and theory seen in  Fig.~\ref{fig:pbpb_b_cms}  is already a non-trivial test of the 
predictive power of the semi-inclusive jet function formalism in QCD matter.  As we pointed  out early on, first predictions for heavy flavor jet substructure using the same in-medium splitting functions have been presented by us~\cite{Li:2017wwc}.  Future measurements of this observable at the LHC and by the sPHENIX experiment at RHIC can further test the universality of the medium induced corrections to b-jets and c-jets.

Figure~\ref{fig:Raa5} presents our predictions for the suppression $R_{\rm AA}$ of c-jets and b-jets (left) and the cross section ratio for different jet radii (right) as a function of jet $p_T$ at $\sqrt{s_{\rm NN}}=5.02$ TeV. Since the CNM effects and collisional energy loss do not depend on the jet radius for  small R, consequently, the radius dependence of $R_{\rm AA}$ is reduced relative to the case where only in-medium radiative processes contribute. For example, in the earlier study of  inclusive light jet modification~\cite{Kang:2017frl}
collisional energy losses were not included and, consequently, the calculated radius dependence was larger.  In p+p collisions,  the ratio of the cross section with $R=0.4$ to that with $R=0.3$ practically does not depend on jet $p_T$.  On the other hand, there is small  dependence on jet $p_T$ in Pb+Pb collisions.  It can be seen from Fig.~\ref{fig:Raa5} that the smaller radius jet tends to dissipate more energy in the medium. There is no significant difference between the c-jet and b-jet due to the high transverse momentum where heavy quark mass effects are small to negligible. In the small transverse momentum region where the uncertainty of our calculations is relatively large,  the contribution from higher orders in opacity~\cite{Sievert:2018imd} might play an important role.  which can be included either by calculating higher order splitting functions~\cite{Fickinger:2013xwa} or by solving the RG equation with the medium-corrected DGLAP kernel \cite{Chien:2015vja}. This can be one of the interesting applications of this framework in the future.

\section{Conclusions} \label{sec:conc}
In this work,  we presented  a  formalism to study heavy flavor jet production in hadronic {\em and} heavy-ion collisions using the heavy flavor semi-inclusive jet functions.  This approach relies on hard-collinear factorization, where the cross section is expressed as the convolution of the PDFs, the hard kernel, and jet functions. For light-flavor jets, similar formalism has been applied to the inclusive jet production and jet substructure yielding gains in the accuracy of theoretical predictions. It has also helped place jet calculations in heavy ion collisions on the firmer theoretical ground. With this in mind, we presented the first calculation of heavy flavor jets in heavy ion collisions using the heavy flavor semi-inclusive jet functions technique.  For jets produced in hadronic collisions, the important $\ln R$ terms were resummed up to leading logarithmic accuracy using a DGLAP-like evolution of the vacuum jet fragmenting functions. In heavy ion collisions the medium corrections are included consistently up to next-to-leading order in QCD and first order in opacity. This limits the applicability of our approach at small 
jet transverse momenta, where such medium corrections can become large and need to also be numerically resummed. We defer this study to future work.

We compared the theoretical cross section results for inclusive c-jet and b-jet production to the experimental data from  p+p and Pb+Pb collisions and found very good agreement between data from the Large Hadron Collider and theory. For the more complex heavy ion collisions, we did include the CNM effects and, for the first time, collisional energy losses in the jet fragmentation function formalism. These were shown to play an important role in the overall suppression of the heavy flavor jet cross sections.  
 We further presented  our calculations of   c-jet and b-jet  cross sections  and their modification  $R_{\rm pA}$ in the cold nuclear matter. 
 Comparison to data at $\sqrt{s_{\rm NN}}$ = 5.02 TeV  demonstrates that, while experimental error bars are large, the CNM effects employed in this calculation are compatible with measurement.  We finally showed the predictions of $R_{\rm AA}$ for the production and attenuation
 of  c-jets and b-jets of different radii  in the highest center-of-mass energy Pb+Pb collisions. With such developments in place,  it 
 might be useful to revisit the calculation of photon-tagged heavy flavor jets~\cite{Huang:2015mva}  and back-to-back b-jets~\cite{Kang:2018wrs},  which so far have only been considered in the  soft gluon approximation medium-induced energy loss limit.

In future,  we plan to further investigate heavy flavor-tagged  jet  substructure observables.  Energy correlators inside jets~\cite{Larkoski:2014gra} are being evaluated in the framework of SCET~\cite{Lee:2019lge}. It would be interesting to calculate them in heavy ion collisions, and for such observables inclusion of  higher orders-in-opacity  corrections in the medium~\cite{Sievert:2018imd} might be important.  We further expect that resummation if the in-medium branchings will improve the predictions in the small-$p_T$ region,  allowing access to the kinematic domain where mass effects on the heavy flavor jet production and propagation in a dense QCD are most pronounced and can lead to novel phenomena~\cite{Li:2017wwc}. 
With a proper extension, we expect that this formalism will be well-suited to investigate jet shapes in hadronic~\cite{Li:2011hy,Chien:2014nsa,Kang:2017mda}  and  heavy ion collisions~\cite{Chien:2015hda}. Other observables of interest are collinear~\cite{Kang:2016ehg}   and transverse momentum fragmentation functions~\cite{Kang:2017glf,Makris:2018npl} for a hadron production inside jets, extended to heavy flavor~\footnote{Note that in Ref. \cite{Anderle:2017cgl} D$^*$ production inside jets was used to constrain fragmentation 
into open heavy flavor,  but the analysis relied on  semi-inclusive fragmenting jet functions for massless partons.}.   The application of the extended formalism to such observables will provide an important test of the universality of the medium modifications predicted by the SCET$_{\rm (M,) G}$ framework.  We finally remark that our approach is also applicable to heavy flavor jet production at a future electron-ion collider.

\section*{Acknowledgements} 
We would like to thank Zhong-Bo Kang, Emanuele Mereghetti and Felix Ringer for useful discussions. 
 This work was supported by the U.S. Department of Energy under Contract No. DE-AC52-06NA25396, its  Early Career Program, and the Los Alamos National Laboratory LDRD program.

\appendix

\bibliographystyle{JHEP}

\bibliography{RAA}

\end{document}